\documentclass[aps,prb,twocolumn,amsmath,amssymb,superscriptaddress,floatfix]{revtex4}
\usepackage{graphicx}
\usepackage{bm}
\usepackage[usenames]{color}
\bibstyle{apsrev.bib}

\newcommand{\be}{\begin{equation}}
\newcommand{\ee}{\end{equation}}
\newcommand{\beqn}{\begin{eqnarray}}
\newcommand{\eeqn}{\end{eqnarray}}

\begin{document}

\title{Renormalization group study of random quantum magnets}
\author{Istv\'an A. Kov\'acs}
\email{ikovacs@szfki.hu}
\affiliation{Department of Physics, Lor\'and E\"otv\"os University, H-1117 Budapest,
P\'azm\'any P. s. 1/A, Hungary}
\affiliation{Research Institute for Solid State Physics and Optics,
H-1525 Budapest, P.O.Box 49, Hungary}
\author{Ferenc Igl\'oi}
\email{igloi@szfki.hu}
\affiliation{Research Institute for Solid State Physics and Optics,
H-1525 Budapest, P.O.Box 49, Hungary}
 \affiliation{Institute of Theoretical Physics,
Szeged University, H-6720 Szeged, Hungary}
\date{\today}

\begin{abstract}
We have developed a very efficient numerical algorithm of the strong disorder renormalization group
method to study the critical behaviour of the random transverse-field Ising model,
which is a prototype of random quantum magnets. With this algorithm we can
renormalize an $N$-site cluster within a time $N \log N$, independently of the topology of the graph
and we went up to $N \sim 4 \times 10^6$.
We have studied regular lattices with dimension $D \le 4$ as well as Erd\H os-R\'enyi
random graphs, which are infinite dimensional objects. In all cases the quantum critical behaviour is found
to be controlled by an infinite disorder fixed point, in which disorder plays a dominant r\^ole over quantum
fluctuations. As a consequence the renormalization procedure as well as the obtained critical properties are
asymptotically exact for large systems. We have also studied Griffiths singularities in the paramagnetic
and the ferromagnetic phases and generalized the numerical algorithm for another random quantum systems.
\end{abstract}


\maketitle
\section{Introduction}
Quantum phase transitions take place at $T=0$ temperature by varying a control
parameter, which is involved in the Hamiltonian of the system\cite{sachdev}. Experimental examples
in which quantum phase transitions play an important role are among others rare-earth magnetic insulators\cite{BRA96}, heavy-fermion compounds \cite{HeavyF,Lohn96}, high-temperature
superconductors\cite{hightc,sciencereview} and two-dimensional electron gases \cite{sondhi,KMBFPD95}. Generally, when quantum fluctuations are weak the ground-state
of these systems is ordered, whereas for strong quantum fluctuations we are in the quantum disordered region. In between
there is a quantum phase-transition point, the effect of which is manifested also in finite temperature in the so called
quantum critical region. 

A paradigmatic model having a quantum phase-transition is the Ising model in the presence of a transverse field of strength, $h$,
where $h$ plays the role of the control parameter. For $h<h_c$ ($h>h_c$) the system is ferromagnetic (paramagnetic) and at
$h=h_c$ there is a quantum critical point. Experimentally this system is realized by the compound ${\rm LiHoF}_4$, which is a
dipole coupled Ising ferromagnet, and which is placed in a magnetic field transverse to the Ising axis of strength, $H_t$.
This magnetic field splits the ground-state doublet and therefore it acts as an effective transverse
field of strength $h \sim H_t^2$. In the above compound one can introduce randomness, by substituting
the magnetic ${\rm Ho}$ by a nonmagnetic ${\rm Y}$. Then
the obtained system ${\rm LiHo}_x{\rm Y}_{1-x}{\rm F}_4$ is the experimental realization of a random quantum magnet\cite{experiment}.
However, the transverse field induces also a random longitudinal field\cite{rf} via the
off-diagonal terms of the dipolar interaction, therefore there are several open questions both theoretically and
experimentally about the low-temperature behavior of this compound.

Here we consider a theoretically simpler problem, the random transverse-field Ising model (RTIM),
which is defined by the Hamiltonian:
\be
{\cal H} =
-\sum_{\langle ij \rangle} J_{ij}\sigma_i^x \sigma_{j}^x-\sum_{i} h_i \sigma_i^z\;.
\label{eq:H}
\ee
Here the $\sigma_{i,j}^{x,z}$ are Pauli-matrices at sites $i$ (or $j$) of a lattice and the
nearest neighbour couplings, $J_{ij}$, and the transverse fields, $h_i$, are independent random numbers,
which are taken from the distributions, $p(J)$ and $q(h)$, respectively.
In this paper we have used two different disorder distributions in order to check universality of the
properties at the critical point. For both disorders the couplings are uniformly distributed:
\be
p(J)=\Theta(J)\Theta(1-J)\;,
\label{uniform}
\ee
where $\Theta(x)$ denotes the Heaviside step-function.

For \underline{box-$h$ disorder} the distribution of the transverse-fields is uniform:
\be
q(h)=\dfrac{1}{h_b}\Theta(h)\Theta(h_b-h)\;,
\label{box-h} 
\ee
whereas for \underline{fixed-$h$ disorder} the initial values of the transverse-fields are constant:
\be
q(h)=\delta(h-h_f)\;.
\label{fix-h} 
\ee
The quantum control parameter is defined as $\theta=\log(h_b)$ and $\theta=\log(h_f)$,
respectively.

Theoretical study of the critical behaviour of the RTIM is a complicated issue, since one should treat the joint
effect of quantum and disorder fluctuations, as well as non-trivial correlations. Most of the results in this
field are known in one dimension (1D), due to a renormalization group (RG) treatment\cite{im}, which has been
introduced by Ma, Dasgupta and Hu\cite{mdh} and further developed by Fisher\cite{fisher}. This RG procedure works in the energy space and
in each step the largest local parameter in the Hamiltonian, either a coupling or a transverse field is decimated
out and at the same time new effective parameters are generated perturbatively. Repeating this procedure
the energy scale in the system, measured by the largest effective parameter goes to zero and one studies
the distributions of the renormalized couplings and that of the transverse fields at the fixed point.

In 1D where the topology of the system is preserved during renormalization Fisher\cite{fisher} has
solved analytically the RG equations and in this way precise information has been obtained about
the behavior of the RG flow in the vicinity of the fixed point. One important result, that the distribution
of the renormalized parameters (couplings and/or transverse fields) at the fixed point are logarithmically broad.
This means that disorder fluctuations are completely dominant over quantum fluctuations and the perturbative treatment
used in the calculation of the new parameters is asymptotically exact. In this, so called infinite disorder
fixed point\cite{danielreview} (IDFP) of the 1D model exact results has been derived
for the critical exponents, as well as for several scaling functions.

In more complex geometries, even for a ladder with $w>1$ legs the topology is changing during the RG process and
therefore one relies on numerical implementations of the RG procedure. For a ladder with a finite width this
procedure can be performed straightforwardly\cite{ladder}, during which the system is renormalized to an effective chain.
As a result the critical singularities of ladders are identical to those of a chain, however, from the $w$-dependence of the amplitudes one can deduce cross-over functions and estimate the singularities of the 2D system through finite size scaling.

If the sample is isotropic, say an $L \times L$ part of a 2D lattice one can not successfully renormalize it numerically with a naive application of
the RG rules. It is due to the fact that after $h$-decimation steps a large number of new effective couplings are generated and
the system will soon look like to a fully connected cluster, for which the further decimations are very slow. To
speed up the process new ideas have been introduced, one of those is the so called maximum rule\cite{motrunich00}.
This is applied in a situation, when between two sites two or more couplings are generated and the maximum of those is taken.
This procedure is certainly correct at an IDFP, where the couplings have typically very different magnitudes.
Together with the maximum rule one can also use some filtering condition\cite{2dRG} to eliminate the latent, non-decimated
couplings of the Hamiltonian, which are at such a position, where a larger parallel coupling will be generated.
Using these tricks one could numerically study the 2D system\cite{motrunich00,lin00,karevski01,lin07,yu07} up to a
linear size $L=128-160$. The
critical exponents calculated in this way are in agreement with the finite-size scaling extrapolations obtained in the
strip geometry\cite{ladder} and with the results of quantum Monte Carlo simulations\cite{pich98}.
The correctness of the method is also checked by comparing results of Monte Carlo simulations about
the random contact process\cite{vojta09}, which is a basic model of reaction-diffusion processes in a random environment. According to an
RG analysis the critical behavior of this model (for strong enough disorder) is identical to that of the RTIM\cite{hiv},
which is indeed found in 1D and 2D calculations.

The model of real random quantum magnets, the RTIM in 3D can not be studied even by the above refined RG algorithm, since
the available finite sizes of the system are too small to obtain stable estimates about the properties of the fixed point.
In early studies the possible presence of an IDFP is expected\cite{motrunich00}, but no evidence in favour of this conjecture has been
presented. Also no studies are available about the random contact process in 3D. In higher dimensions no results of any
kind are known, thus it is a completely open question, if there is an upper critical dimension, $D_u$, such that for
$D<D_u$ infinite disorder scaling works and for $D \ge D_u$ we have conventional random criticality. Furthermore, it is
also an open question, if for $D \ge 2$ the fixed point is universal and does not depend on the actual form of the
disorder.

A possible way to answer to the questions presented in the previous paragraph is to improve the numerical algorithm
of the RG procedure and make it capable to study three- and higher dimensional systems. In this paper we describe such
an improved algorithm, which uses the maximum rule but otherwise has no further approximations. This means that the
results of our algorithm are identical to that of any naive implementation of the RG method (having also the maximum
rule) for any finite graphs, say with $N$ sites and $E$ edges. However, we gain considerable time in performance:
while the naive method works in $t \sim \mathcal{O}(N^3)$ time, this is for the improved algorithm
$t \sim \mathcal{O}(N \log N + E)$. Having this performance at hand we could treat finite clusters up to
$4 \times 10^6$ sites.

We have used this algorithm to study the properties of the RTIM in different dimensions. Some preliminary results
of these investigations have already been presented elsewhere. The critical behaviour of the 3D and 4D systems
is announced in\cite{34DRG}. Here we give details of the determination of the critical parameters as well as
analyze the scaling behavior of different quantities in the off-critical region, too. The model in 2D
has been studied by another algorithm in\cite{2dRG}, which algorithm has basically
the same performance as the present one in 2D, however, which is less effective in higher dimensions.
Here the 2D results are merely used to compare those with higher dimensional results.
We have also studied
Erd\H os-R\'enyi random graphs\cite{erdos_renyi} with a finite coordination number, which are infinite dimensional objects. In this way we
have got information about the possible value of the upper critical dimension, $D_u$, in our system.

Our paper is organized as follows. The RG procedure, the basic decimation rules and the essence of the improved
algorithm are presented in Sec.\ref{sec:RG}. Results about the critical behaviour of the system in different dimensions
are presented in Sec.\ref{sec:results}. The results are discussed in Sec.\ref{sec:discussion} and possible extension of the improved algorithm for another models are given in the Appendix.

\section{The renormalization group method and the improved algorithm}
\label{sec:RG}

The so called strong disorder RG (SDRG) method\cite{im} has been introduced by Ma and coo-workers\cite{mdh} to study 1D random
antiferromagnetic Heisenberg chains. For the RTIM it was Fisher\cite{fisher}, who used first this method and solved analytically the
RG equations in 1D. Here we discuss the basic steps of the RG procedure for the RTIM model.

The SDRG method works in the energy space: at each step the largest parameter of the Hamiltonian in
Eq.(\ref{eq:H}), which is denoted by $\Omega=\max\{J_{ij},h_i\}$, is eliminated. Here we should consider two possibilities.

\underline{$J$-decimation}:~
If the largest term is a coupling, say $\Omega=J_{ij}$, than the two connected sites $i$ and $j$ are merged into a spin cluster
having an effective moment $\tilde{\mu}=\mu_i+\mu_j$ (in the original model $\mu_i=1, \forall i$), which is
placed in an effective transverse field of strength: $\tilde{h}=h_i h_j/J_{ij}$. This latter formula is obtained
in second-order perturbation calculation.

\underline{$h$-decimation}:~
If the largest term is a transverse field, say $\Omega=h_i$, than this site brings a negligible contribution to the
(longitudinal) susceptibility of the system and therefore decimated out. At the same time new couplings are generated
between all sites, say $j$ and $k$, which were nearest neighbours to $i$. In second-order perturbation calculation
these couplings are given by $J_{ji}J_{ik}/h_i$. If there is already a coupling, $J_{jk}>0$, between
the two sites we use the maximum rule: $\tilde{J}_{jk}=\max\{J_{ji}J_{ik}/h_i,J_{jk}\}$.

In the naive application of the SDRG rules for higher dimensional clusters there is a problem with the
$h$-decimation steps, during which several new couplings are generated. As a result our cluster will be transformed
soon into an almost fully connected graph, having $\mathcal{O}(N^2)$ edges. Consequently at any further decimation
step one needs to perform $\mathcal{O}(N^2)$ operations, which leads to a performance in time $t \sim \mathcal{O}(N^3)$.

\subsection{Improved algorithm}
\label{sec:improved}

Here and in the following we assume, without restricting generality, that $J_{ij}\leq 1, \forall i,j$.
Then we define the set of \textit{local maximum}, which consists of such parameters, which are larger
(not smaller) than any of its neighbouring terms.  Considering a coupling $J_{ij}$ is a local maximum, provided $J_{ij} \ge h_i$, $J_{ij} \ge h_j$, and
$J_{ij}\ge J_{ik},~\forall k$, as well as $J_{ij}\ge J_{lj},~\forall l$. Similarly a transverse field, $h_i$,
is a local maximum, if $h_i \ge J_{ij},~\forall j$. We have shown\cite{2dRG} that the local maximum can be decimated independently,
the renormalization performed in any sequence gives the same final result. 

In the improved algorithm we concentrate on the transverse fields, the decimation of which being the most
dangerous in respect of the performance of the algorithm. Our strategy is to avoid any $h$-decimation
during the renormalization. For this purpose we divide the sites into two classes.
For 'inactive' sites the transverse field is a local maximum and the site has a weight, $l_i=1$,
whereas all the remaining sites are termed as 'active' having weights, $l_i=0$. In the log-energy space between sites $i$ and
$j$ we define a distance, $d_{ij}\geq 0$, as:
\be
d_{ij} = -\ln{J_{ij}}+\frac{l_i}{2}\ln{h_{i}}+\frac{l_j}{2}\ln{h_{j}}.
\label{eq:d}
\ee
Having an inactive site, $k$, between $i$ and $j$ and decimating it out the RG rules lead to the additivity property: $\tilde{d}_{ij}=d_{ik}+d_{kj}$, which - according to the maximum rule - should be compared with $d_{ij}$ in Eq.(\ref{eq:d})
and their minimal value is taken. Generally, the true distance between $i$ and $j$, denoted by $\delta_{ij}$
is given by the shortest path which goes over the inactive sites.
It is also easy to see, that decimating out all or a subset of inactive sites is equivalent to find in the original problem the shortest paths among the non-decimated sites which go
through the decimated sites. This is a well known graph-theoretical problem\cite{dijkstra} for which efficients numerical
algorithms are available.\cite{spanning_tree} In general, however, one should also deal with the active sites, and
therefore $J$ decimations should also be performed.

In the following, we concentrate on the active sites, and define to each in the
log-energy space a range, 
\be
r_i=-\ln{h_i}.
\label{eq:r}
\ee
In the improved algorithm we compare the ranges of the
active sites with the true distances measured between them. In this respect two possibilities may happen.

\begin{itemize}
\item If the true distance between two active sites ($i$ and $j$) is not greater than any of their ranges, i.e.
$\delta_{ij}\leq{r_i}$ and $\delta_{ij}\leq{r_j}$, then $i$ and $j$ are fused together into an effective active site,
which has a range 
\be
\tilde{r}=r_i+r_j-\delta_{ij}.
\label{eq:rupdate}
\ee
Also the distance measured from this effective site to another site, say $k$, is given by
$\min(d_{ik},d_{jk})$.
\item If the range, $r_i$, of an active site, $i$ is shorter than any of its true distances from active sites,
$r_i<\delta_{i,j},  \forall j$, then this site can not be fused together with any other
active sites, therefore it is turned to 'inactive'.
Then we set its weight $l_i=1$ and update the distances, $d_{ij}, \forall j$, in Eq.(\ref{eq:d}). 
\end{itemize}

In the above renormalization steps, which can be used in arbitrary order, the number of active sites in the system
is reduced by one.
Repeating these decimation rules we arrive to a system having only inactive sites and no further fusion steps take place.
The complete cluster-structure including the excitation energies are readily encoded in this configuration,
which can be extracted without further renormalization steps.
In any case the final result of the improved algorithm is identical to that obtained by the na\"{\i}ve SDRG algorithm.
\begin{figure}[h!]
\begin{center}
\includegraphics[width=3.4in,angle=0]{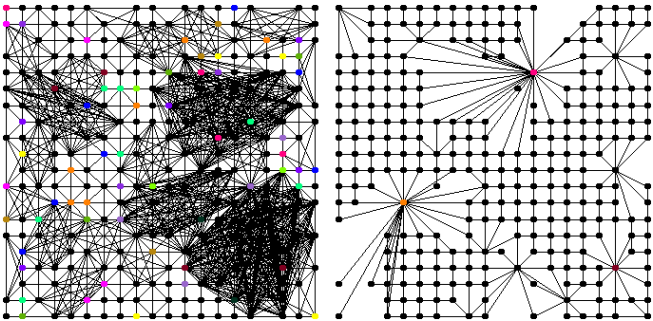}
\end{center}
\vskip -.5cm
\caption{
\label{fig_1} (Color online)
Snapshots of the na\"{\i}ve (left) and improved (right) SDRG algorithms for the same critical $20 \times 20$ sample
with box-$h$ disorder and open boundary conditions, where $40\%$ of the spins is decimated out. While
in the na\"{\i}ve algorithm numerous new couplings are generated in the improved algorithm we merely delete sites.
}
\end{figure}

In practice one should measure the true distances between the active sites. This can be done from a selected
reference site from which we measure the distances by Dijkstra's method\cite{dijkstra}
including one nearby site after the other until the range, $r_i$, is reached. While repeating the exploration of the
shortest paths from
all active sites, a given site could be crossed from several directions. However, it can be shown, that it is always sufficient to cross it from one direction only\cite{34DRG}. As a consequence, we can delete those sites, which have
already been explored, from the system, because these are no longer needed for the calculation. 
Therefore in a more efficient algorithm initially we consider all active sites and perform the measurements simultaneously
and successively delete the explored sites.
The implementation of this algorithm has a time complexity of $\mathcal{O}(N \log N + E)$ on any graphs with $N$ sites and $E$ edges. A further speed gain is achieved by recognizing, that in this parallel implementation the paths only need to be explored until reaching a length of $r_i/2$ instead of the full $r_i$ range. In Fig.\ref{fig_1}
we compare the topology of the renormalized systems in the na\"{\i}ve and improved SDRG algorithms.
\begin{figure}[h!]
\begin{center}
\includegraphics[width=3.4in,angle=0]{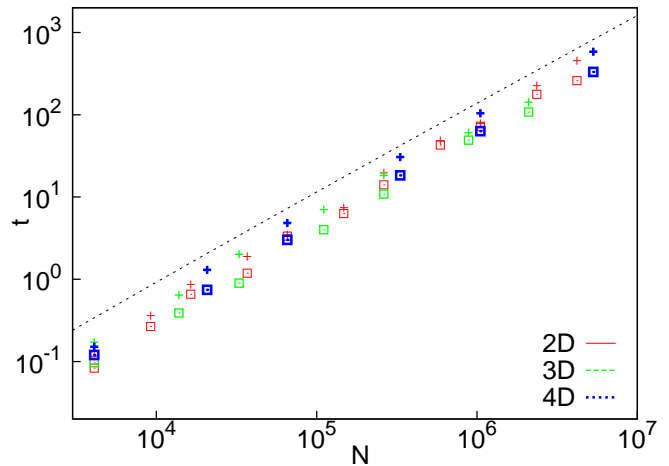}
\end{center}
\vskip -.5cm
\caption{
\label{fig_2} (Color online)
Computational time of the algorithm, $t$, as a function of the size of the hypercubic clusters, $N$,
in a log-log scale for 2D, 3D and 4D and for the two different disorders (fixed-$h$ +, box-$h$ $\boxdot$).
The theoretical prediction, $t \sim N \log N$, is indicated by a dashed line.
}
\end{figure}

We have checked the computational time of the algorithm, $t$, for 2D, 3D and 4D hypercubic clusters
consisting of $N$ sites. The results are shown in Fig.\ref{fig_2} for the two types of randomness.
For a given $N$ and for a given type of disorder the computation time is
practically independent of the topology of the cluster and $t$ is well described by the theoretical bound: $\sim  N \log N$.
(In the studied cases the number of edges are proportional to $N$.) Generally, for a given $N$ the renormalization
for the box-$h$ randomness is faster.

\section{Critical behaviour in different dimensions}
\label{sec:results}
\subsection{Analytical results in 1D - a reminder}
In 1D the position of the critical point is given by the condition\cite{pfeuty}: $\delta=[ \ln h]_{\rm av}-[ \ln J]_{\rm av}=0$,
where $[ \dots ]_{\rm av}$ stands for averaging over quenched disorder.
At the critical point the energy-scale, $\Omega$, and the length-scale, $L$, which is the linear size of the system are
related as: $\log(\Omega_0/\Omega) \sim L^{\psi}$, with $\psi=1/2$. This type of unusual dynamical scaling relation is
a clear signal of infinite disorder scaling.
In the paramagnetic phase, $\delta>0$, the spin clusters have a finite extent, $\xi$, which in the vicinity
of the critical point diverges as $\xi \sim \delta^{-\nu}$, with $\nu=2$. At the
critical point the largest cluster is a fractal, having a moment: $\mu \sim L^{d_f}$, with a fractal dimension:
$d_f=(1+\sqrt{5})/4$. From this the magnetization exponent is expressed by $\beta=x \nu$, with $x=d-d_f$ and $d=1$.
These SDRG results\cite{fisher}, which have been extended to the dynamical properties in the Griffiths phases\cite{igloi02} have been tested by independent analytical\cite{mccoywu,shankar} and numerical calculations\cite{young_rieger96,igloi_rieger97,bigpaper}.

\subsection{Numerical results for $D=3$ and $4$}
\label{sec:234d}
We have studied finite systems of hypercubic lattices in dimensions $D=3$ and $4$, the largest
linear sizes of the samples were $L=128$ and $48$, respectively. (In 2D the previously performed
numerical investigations\cite{2dRG} went up to $L=2048$.) For each sizes we have renormalized typically $40000$
random samples (for each type of disorder), but even for the largest systems we have treated at least
$10000$ realizations.

\subsubsection{Finite-size critical points}

The precise identification of the critical point of a disordered system is a very important issue, since the accuracy
of the determination of the critical exponents depends very much on it. This question is deeply related to the
problem of finite-size scaling in random systems\cite{psz,domany,aharony,paz2,mai04,PS2005}, since estimates for the critical points are generally calculated
in finite samples. As has been observed recently\cite{domany,aharony} the key issue in this point is the
scaling behavior of the finite-size critical points, which are calculated for a large set of samples. For a given
sample, denoted by $\alpha$, this pseudo-critical point, $\theta_c(L,\alpha)$, is located at the point, where
some physical parameter of the system has a maximal value. For example in a classical random magnet the susceptibility
can be used for this purpose, which is divergent at the true critical point of the infinite system.

For a random
quantum system the susceptibility is divergent in a whole region, in the so called Griffiths phase, thus
can not be used to monitor finite-size critical points. Then in 1D random quantum systems the average entanglement
entropy turned out to be a convenient quantity, the maximum of which is a good indicator of finite-size
criticality\cite{ilrm}. In more complicated topology, for ladders\cite{ladder} and in 2D systems\cite{2dRG} 
the so called doubling method\cite{PS2005} is found to provide an appropriate definition of $\theta_c(L,\alpha)$.
In this procedure one considers two identical copies of a given sample, $\alpha$, which are joined together by surface couplings
and this replicated sample is denoted by $2\alpha$.
Using the SDRG method one calculates some physical quantity (magnetization or gap)
in the original and in the replicated sample, which are denoted by $f(\alpha)$ and $f(2\alpha)$, respectively,
and study their ratio, $r(\alpha)=f(2\alpha)/f(\alpha)$, as a function of the control parameter, $\theta$.
At $\theta=\theta_c(\alpha,L)$ this ratio has a sudden jump, which is identified with the pseudo-critical point of
the sample. It has been realized\cite{2dRG} that this jump in the ratio is related to a sudden change in the
cluster structure which is generated during the SDRG procedure. For weak
quantum fluctuations, $\theta<\theta_c(L,\alpha)$, between the replicas correlations are generated during renormalization,
which are manifested
by the presence of a so called \textit{correlation cluster}. This contains equivalent sites in the two replicas. On the contrary for stronger quantum fluctuations, $\theta>\theta_c(L,\alpha)$,
the two replicas are renormalized independently. For $\theta<\theta_c(L,\alpha)$ the mass of the correlation cluster, $\mu(L,\alpha,\theta)$, is a monotonously
decreasing function of $\theta$. Then we identify
$\theta_c(L,\alpha)$ as the point where the correlation cluster disappears.

\begin{figure}[h!]
\begin{center}
\includegraphics[width=3.4in,angle=0]{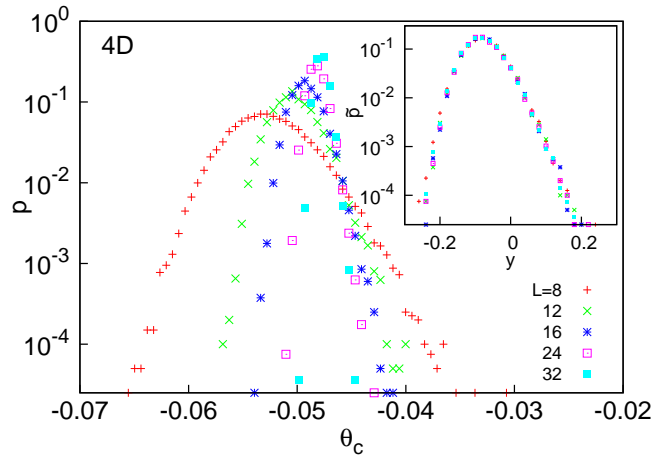}
\end{center}
\vskip -.5cm
\caption{
\label{fig_3} (Color online)
Distribution of the pseudo-critical points, $\theta_c(L)$, for various sizes for fixed-$h$ randomness
for 4D. In the inset the scaled distributions are shown as a function of
$y=(\theta_c(L)-\theta_c)L^{1/\nu}$, see the text.
}
\end{figure}

\begin{figure}[h!]
\begin{center}
\includegraphics[width=3.4in,angle=0]{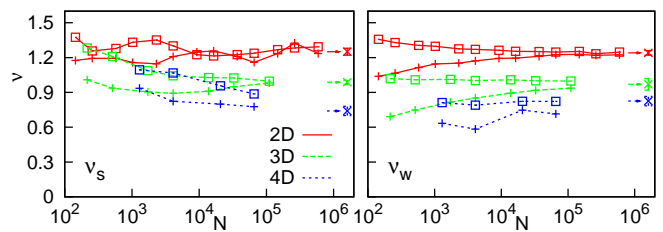}
\end{center}
\vskip -.5cm
\caption{
\label{fig_4} (Color online)
Finite-size estimates for the shift, $\nu_s$, (left) and the width, $\nu_w$, (right)
critical exponents for 2D, 3D and 4D and for the two different disorders (fixed-$h$ +, box-$h$ $\boxdot$).
The estimated values, as given in Table.\ref{table:1} are indicated at the right edge of the figures.
}
\end{figure}

Using this doubling method we have calculated pseudo-critical points in different dimensions and for
different forms of the disorder. We illustrate the distributions of the pseudo-critical points in Fig.\ref{fig_3}
for $D=4$. To analyze these distributions we make use results of finite-size scaling theory\cite{domany,aharony},
which makes statements about the average value, $\overline{\theta}_c(L)$, and the width of the distribution
, $\Delta \theta_c(L)$, in finite systems of linear size, $L$. The average value of the distribution
is expected to scale as:
\be
\left|\theta_c-\overline{\theta_c}(L)\right| \sim L^{-1/\nu_s}\;,
\label{nu_s} 
\ee
where $\theta_c$ is the true critical point of the system and $\nu_s$ is the so called shift exponent.
On the other hand the width of the distribution is expected to scale as:
\be
\Delta \theta_c(L) \sim L^{-1/\nu_w}\;,
\label{nu_w} 
\ee
with a width exponent, $\nu_w$.
We have checked that these
finite-size scaling relations are satisfied in all dimensions, and these
relations are used - by comparing results at two different sizes (at $L$ and $L/2$) - to obtain finite-size
estimates for the exponents. These are presented in Fig.\ref{fig_4} for the two different forms of disorder. 

For a given
dimension the estimated exponents are found to be independent of the form of the disorder, furthermore
the shift and the width exponents are identical within the error of the calculation. Our estimates about the critical exponents
are collected in Table.\ref{table:1}, together with the estimates of the true critical points.
We note, that the relation $\nu_s=\nu_w=\nu$ is characteristic for scaling at a conventional random
fixed point, and the distributions of the pseudo-critical points can be rescaled to a master curve in terms
of the variable, $y=(\theta_c(L)-\theta_c)L^{1/\nu}$, which is shown in the inset of Fig.\ref{fig_3}.

\begin{table}
\caption{Critical properties of the RTIM in different dimensions. In 1D the analytical
results are from\cite{fisher}, in 2D the numerical results are taken from\cite{2dRG}.\label{table:1}}
 \begin{tabular}{|c|c|c|c|c|}  \hline
   & 1D & 2D & 3D & 4D\\ \hline
$\theta_c^{(b)}$   &  $0$ & $1.6784(1)$ & $2.5305(10)$ & $3.110(5)$  \\ 
$\theta_c^{(f)}$   &  $-1.$ & $-0.17034(2)$ & $-0.07627(2)$ & $-0.04698(10)$  \\ \hline
$\nu_w$   &  $2.$ & $1.24(2)$ & $0.97(5)$ & $0.825(40)$  \\ 
$\nu_s$   &   & $1.25(3)$ & $0.987(17)$ & $0.74(4)$  \\ \hline
$x$   &  $\frac{3-\sqrt{5}}{4}$ & $0.982(15)$ & $1.840(15)$ & $2.72(12)$  \\ \hline
$\psi$   &  $1/2$ & $0.48(2)$ & $0.46(2)$ & $0.46(2)$  \\ \hline
  \end{tabular}
  \end{table}

\begin{figure}[h!]
\begin{center}
\includegraphics[width=3.4in,angle=0]{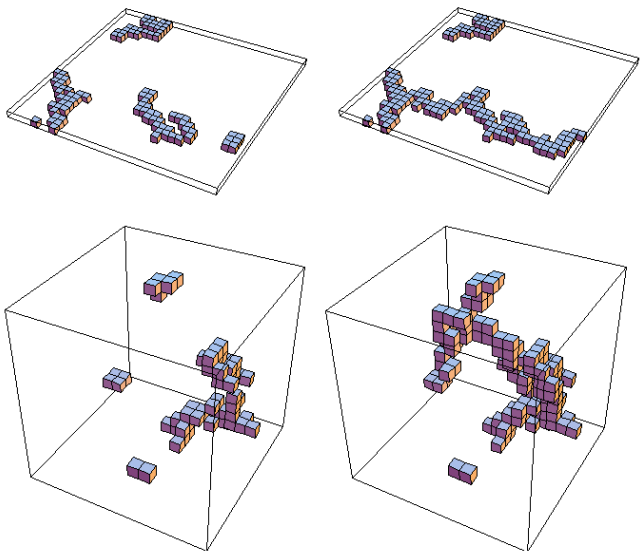}
\end{center}
\vskip -.5cm
\caption{
\label{fig_5} (Color online)
Left panel: correlation clusters at the critical point for fixed-$h$ randomness in 2D ($L=32$) and 3D ($L=16$).
Right panel: the connected subgraphs, which contain the correlation clusters (see text).}
\end{figure}

\subsubsection{Magnetization}

\begin{figure}[h!]
\begin{center}
\includegraphics[width=3.4in,angle=0]{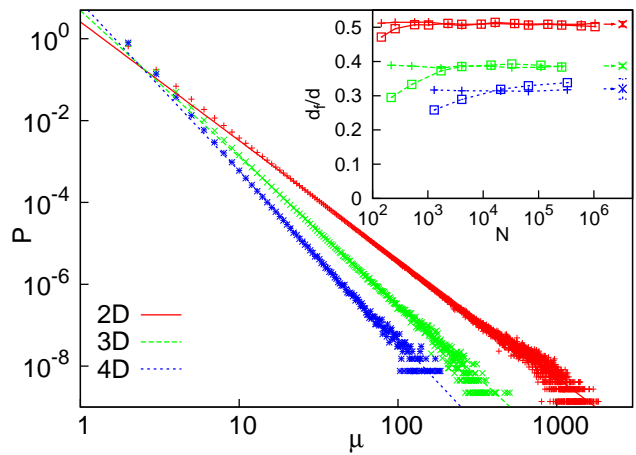}
\end{center}
\vskip -.5cm
\caption{
\label{fig_6} (Color online)
Distribution of the mass of the clusters, $P_L(\mu)$, at the critical point
for 2D ($L=1024$), 3D ($L=128$) and 4D ($L=48$) with box-$h$ randomness
in a log-log scale. The scaling results about the asymptotic slopes of the curves are indicated by straight lines.
Inset: Finite-size estimates for the fractal dimension of the critical correlation cluster for 2D, 3D and 4D and
for the two different disorders (fixed-$h$ +, box-$h$ $\boxdot$).
The estimated values, as given in Table.\ref{table:1} are indicated at the right edge of the figure.}
\end{figure}

The magnetization of the system, $m(L,\theta)$, is related to the average mass of the largest effective clusters,
$\overline{\mu}(L,\theta)$, as $m(L,\theta)=\overline{\mu}(L,\theta)/L^d$. In the thermodynamic
limit, $L \to \infty$, in the ferromagnetic phase, $\delta=\theta-\theta_c<0$ the magnetization is finite and
vanishes at the critical point as:
$\lim_{L \to \infty} m(L,\theta) \sim (-\delta)^{\beta}$ where $\beta$ is the
magnetization exponent. At the critical point the largest connected clusters are fractals, which are illustrated
in the left panel of Fig.\ref{fig_5} for 2D and 3D. The mass of these critical
clusters scales as $\mu \sim L^{d_f}$, where $d_f$ is the appropriate fractal dimension
which is related to the anomalous
dimension of the magnetization as $x=\beta/\nu=d-d_f$. Comparing the average mass
of the largest clusters at two finite sizes, we have calculated effective, size-dependent fractal dimensions,
as well as effective magnetization scaling dimensions. These are presented in the inset of Fig.\ref{fig_6} for the different dimensions
and using the two disorder distributions in Eqs.(\ref{box-h}) and (\ref{fix-h}). Extrapolating these values the obtained
exponents, for a given dimension, do not depend on the form of the disorder. These are presented in Table.\ref{table:1}.

We have also studied the distribution function of the mass of the clusters, $P_L(\mu)$, which is expected
to behave at the critical point as: $P_L(\mu)=L^{d_f} \tilde{P}(\mu L^{-d_f})$.  According
to scaling theory\cite{perco} $\tilde{P}(u)$ for large arguments has a power-law tail, $\tilde{P}(u) \sim u^{-\tau}$,
with $\tau=1+\dfrac{d}{d_f}$. In Fig.\ref{fig_6} we have plotted $P_L(\mu)$ for $D=2,3$ and $4$ for $L=1024,128$ and $48$,
respectively, and good agreement with scaling theory is found.

\begin{figure}[h!]
\begin{center}
\includegraphics[width=3.4in,angle=0]{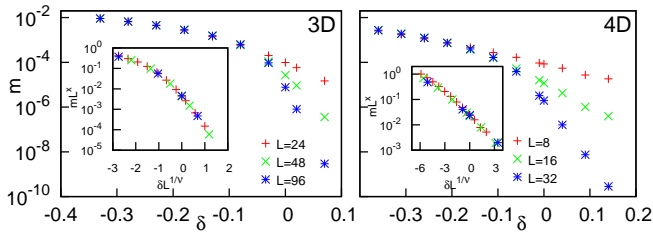}
\end{center}
\vskip -.5cm
\caption{
\label{fig_7} (Color online)
Finite-size magnetizations in the vicinity of the critical point for the 3D (left) and the 4D (right)
models for box-$h$ randomness. In the insets the scaled magnetizations are presented, see the text.}
\end{figure}

Close to the critical point the finite-size magnetizations are shown in Fig.\ref{fig_7} for 3D and 4D.
For large $L$ in the ordered phases ($\delta < 0$) the magnetization approaches a finite limiting value, whereas for $\delta>0$ it
tends to zero. In the vicinity of the critical point the finite-size magnatizations can be transformed to a master curve, if
one considers the scaled magnetization, $\tilde{m}=mL^x$, as a
function of the scaling variable, $\tilde{\delta}=\delta L^{1/\nu}$. This is illustrated in the insets of Fig.\ref{fig_7}
where the exponents $x$ and $\nu$ are taken from Table.\ref{table:1}.

\subsubsection{Dynamical scaling}

\begin{figure}[h!]
\begin{center}
\includegraphics[width=3.4in,angle=0]{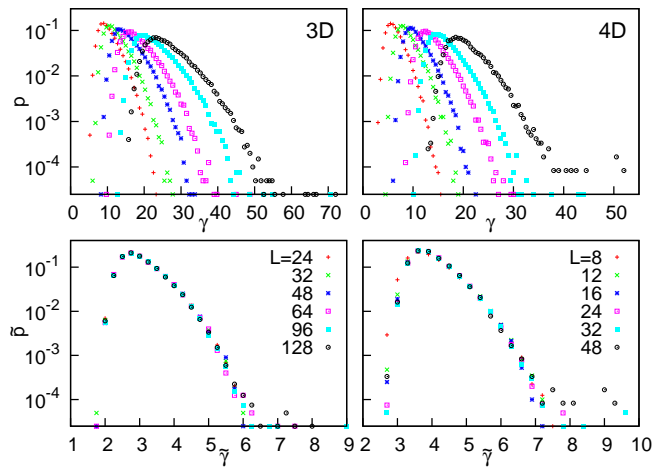}
\end{center}
\vskip -.5cm
\caption{
\label{fig_8} (Color online)
Distribution of the log-energy parameters as the system size in 3D and 4D for box-$h$ randomness at the critical point (upper panel). The scaled distributions are shown in the lower panel, using the
non-universal parameters: $\gamma_0=3.1$ (3D) and $\gamma_0=4.06$ (4D), as described in the text.}
\end{figure}

The dynamical behavior of the RTIM is related to the low-energy excitations of the system,
the energy of which in the SDRG method is given by the values of the effective transverse
fields at which a given cluster is eliminated. Generally, for each such eliminated
cluster one can define a connected subgraph, which contains the given cluster and
renormalization of this subgraph gives the same energy value. The form of the connected
subgraphs of the correlation clusters is illustrated\cite{subgraph} in the right panel of Fig.\ref{fig_5}.
The energy-parameter of a given sample, which is denoted by $\epsilon(L,\alpha)$,
is given by the smallest effective transverse field, not considering the transverse field of the correlation
cluster, if it exists in the system. The distribution of the log-energy parameter,
$\gamma(L,\alpha)=-\log \epsilon(L,\alpha)$, is shown in the upper panel of Fig.\ref{fig_8} at the critical points
of the 3D and 4D systems. The distributions for both dimensions are broadening with increasing $L$, which is a clear
signature of infinite disorder scaling. The typical value of the log-energy parameter
grows with the size as $\gamma(L) \sim L^{\psi}$, thus the appropriate scaling combination is given by:
$\tilde{\gamma}=(\gamma(L)-\gamma_0) L^{-\psi}$. Here $\psi$ is a scaling exponent and $\gamma_0$ is a non-universal
constant. The scaled distributions are shown in the lower panel of Fig.\ref{fig_8}.

\begin{figure}[h!]
\begin{center}
\includegraphics[width=3.4in,angle=0]{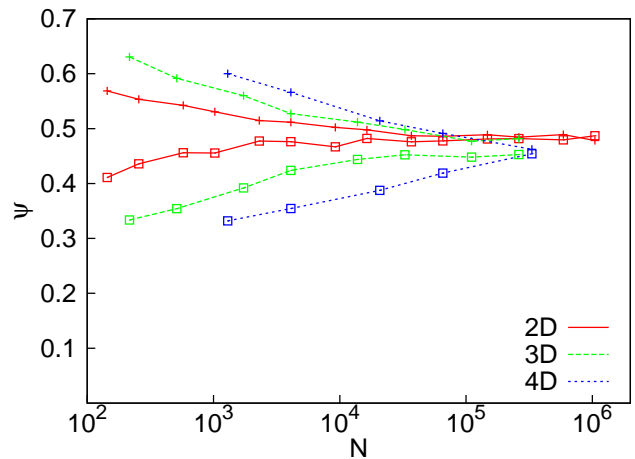}
\end{center}
\vskip -.5cm
\caption{
\label{fig_9} (Color online)
Finite-size estimates for the $\psi$ critical exponent for 2D, 3D and 4D obtained from the width of the log-energy distributions for the two different disorders (fixed-$h$ +, box-$h$ $\boxdot$).}
\end{figure}

We have calculated effective, size-dependent $\psi$ exponents, by comparing the widths of the distributions of the
log-energy parameters at two sizes ($L$ and $L/2$). These are given in Fig.\ref{fig_9} for the different dimensions
and for the two different form of disorder. As for other exponents studied before, the estimates for $\psi$
for a given dimension do not depend on the actual form of the randomness. These are summarized in Table.\ref{table:1}.
Interestingly the $\psi$ exponents for all studied dimensions are close to $1/2$, which is the exact value in 1D. This
observation can be explained by the fact, that the connected subgraphs, which are related to the energy-parameter of the sample,
are basically one-dimensional objects in all studied dimensions, see in the right panel of Fig.\ref{fig_5}. The very
small variation of $\psi$ with the dimensionality is probably due to the fact that the renormalized couplings and
transverse fields of the connected subgraphs are more and more correlated in higher dimensions.

Finally we note that thermodynamic singularities at a small temperature, $T$, but $\delta=0$ are related to the critical exponents
in Table.\ref{table:1}. For example the susceptibility, $\chi$, and the specific heat, $C_V$, behave as\cite{danielreview,im}:
$\chi(T)\sim (\log T)^{(d-2x)/\psi}/T$ and $C_V(T) \sim (\log T)^{-d/\psi}$. The similar relations
at $T=0$ but with a small longitudinal field, $H$, are given by:
$\chi(H)\sim (\log H)^{-x/\psi}/H$ and $C_V(H) \sim (\log H)^{-d/\psi}$.

\subsubsection{Griffiths effects}

\begin{figure}[h!]
\begin{center}
\includegraphics[width=3.4in,angle=0]{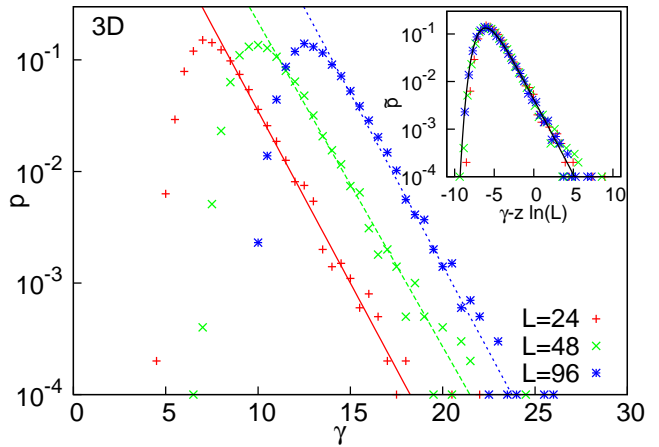}
\end{center}
\vskip -.5cm
\caption{
\label{fig_10} (Color online)
Distribution of the log-excitation energies in the disordered
Griffiths-phase of the 3D model for box-$h$ disorder at $\delta=0.37$
in a log-lin scale for different sizes. The slopes of the straight
lines indicating the tail of the curves are
$d/z=0.71(2), 0.69(2)$ and $0.71(2)$, for $L=24,48$ and $96$,
respectively. In the inset the scaled distributions
are shown with $d/z=0.73$, which is well described by the Fr\'echet
distribution with $\gamma_0=6.09$ (full line).}
\end{figure}

\begin{figure}[h!]
\begin{center}
\includegraphics[width=3.4in,angle=0]{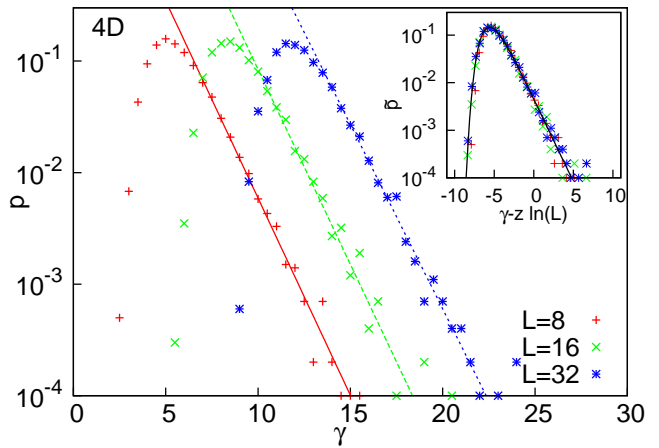}
\end{center}
\vskip -.5cm
\caption{
\label{fig_11} (Color online)
The same as in Fig.\ref{fig_10} for 4D at $\delta=0.34$. The slopes of the straight
lines indicating the tail of the curves are
$d/z=0.81(3), 0.81(3)$ and $0.76(3)$, for $L=8,16$ and $32$,
respectively. In the inset the scaled distributions
are shown with $d/z=0.8$, and the full line represents the Fr\'echet
distribution with $\gamma_0=5.60$.}
\end{figure}

We have also studied the distribution of the low-energy excitations outside the critical point. In the paramagnetic phase
the distribution of $\gamma(L)$ for different sizes are shown in Fig.\ref{fig_10} and in Fig.\ref{fig_11},
for the 3D model ($\delta=0.37$) and for the 4D model ($\delta=0.34$), respectively.
As seen in these figures the distributions have approximately the same width, they are merely shifted with
increasing $L$. This behaviour is in agreement with scaling theory in the disordered Griffiths-phase\cite{im}, where
the typical excitation energy scales with the size as: $\epsilon(L) \sim L^{-z}$, where $z=z(\delta)$ is the
dynamical exponent, which depends on the distance from the critical point, $\delta$. Then the appropriate
scaling combination is: $\tilde{\gamma}=\gamma(L)-z \log L -\gamma_0$, in terms of which
a scaling collapse of the distributions are found, which is shown in the inset of Fig.\ref{fig_10} and Fig.\ref{fig_11}. One way
to estimate $z(\delta)$ is to analyze the scaling collapse of the distributions, or equivalently to compare the
shift of the distributions with $L$. From this type of analysis we obtain $d/z=0.75(4)$ for the 3D
model and $d/z=0.80(4)$ for 4D.
There is, however, another possibility to calculate $d/z$ from the asymptotic form of the distributions. If the
low-energy excitations are localised, which is satisfied for the RTIM, the distribution function of the scaled
variable, $\tilde{\gamma}$, is expected to follow extreme value statistics\cite{jli06} and given by the
Fr\'echet distribution\cite{galambos}:
\be
\ln p(\tilde{\gamma}-\gamma_0)=-\dfrac{d}{z}\tilde{\gamma}-\exp\left( -\tilde{\gamma}\dfrac{d}{z}\right) +
\ln\left(\dfrac{d}{z} \right)\;.
\label{frechet}
\ee
Indeed the scaled distributions in the insets of Figs.\ref{fig_10} and \ref{fig_11} are well described by this form,
having just one free parameter, $\gamma_0$. From Eq.(\ref{frechet}) follows that the asymptotic slope of
$\ln p(\gamma)$ vs. $\gamma$ is just $d/z$, what we have measured in Figs.\ref{fig_10} and \ref{fig_11}. The estimates for
$d/z$ for different values of $L$, as given in the captions are in good agreement with our
previous estimates from the shift of the curves.

We have repeated these type of calculations at other points of the disordered Griffiths phase and we have
measured $\delta$ dependent dynamical exponents. We could not, however, check the scaling
result\cite{im,danielreview} for small $\delta$: $d/z \propto \delta^{\nu \psi}$, due to strong finite-size
effects in the vicinity of the critical point. (In 2D this type of analysis has been performed in\cite{2dRG}.)

\begin{figure}[h!]
\begin{center}
\includegraphics[width=3.4in,angle=0]{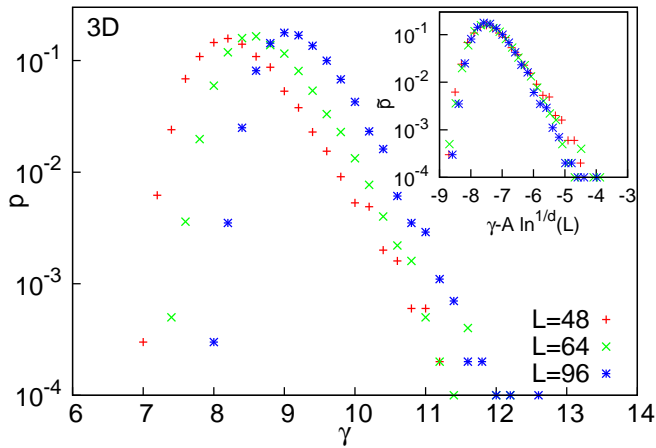}
\end{center}
\vskip -.5cm
\caption{
\label{fig_12} (Color online)
The same as in Fig.\ref{fig_10} for 3D at the ordered Griffiths phase, $\delta=-0.33$.
In the inset the scaled distributions are shown with $A=10.0$.}
\end{figure}

We have also calculated the distribution of the log-energy parameters in the ordered Griffiths phase,
which is illustrated in Fig.\ref{fig_12} for the 3D model at $\delta=-0.33$. In the ordered phase there is a huge
magnetization cluster, which has a very small effective field and the energy parameter is given by the
second smallest effective field of the RG process. As in the disordered Griffiths phase the width of
the distributions is approximately $L$ independent and the distributions are shifted with $L$. However, the
amount of shift with $L$, as well as the shape of the distributions are different in the two cases. This
is related to the scaling result, that the typical value of the excitation energy in the ordered Griffiths phase
scales with the size as: $\epsilon(L) \sim - \ln^{1/d}(L)$, thus the appropriate scaling combination is:
$\tilde{\gamma}=\gamma(L)-A \ln^{1/d}(L)-\gamma_0$, with $A$ and $\gamma_0$ being nonuniversal constants.
Using this variable the distributions have a scaling collapse as shown in the inset of Fig.\ref{fig_12}.
We should, however, mention that due to finite-size effects we can not obtain an independent
estimate of the exponent of the logarithm, being theoretically $1/d$. We are facing to the same kind of
limitations concerning the shape of
the scaling curve, which asymptotically should behave as: $\ln p(\tilde{\gamma}) \sim -\tilde{\gamma}^d$,
according to scaling theory. Our data, however, are still not in the asymptotic regime.

Finally we note that  in the disordered Griffiths phase the singularities of the susceptibility and the
specific heat at a small temperature are
given by\cite{motrunich00,im}: $\chi(T) \sim T^{-1+d/z}$ and $C_V(T) \sim T^{d/z}$. The same expressions in the ordered Griffiths
phase are: $\chi(T) \sim \exp(-C |\log T|^d)/T$ and $C_V(T) \sim \exp(-C' |\log T|^d)$.

\subsection{Numerical results for Erd\H os-R\'enyi random graphs}

Here we come back to the question posed in the Introduction about the possible value of the upper critical
dimension, $D_u$, in the problem. In order to answer to this question we consider
Erd\H os-R\'enyi (ER) random graphs\cite{erdos_renyi} with a finite coordination
number, which are representing the large-dimensional limit of our lattices. 
Generally an ER random graph consists of $N$ sites and $kN/2$ edges, which are put in random
positions. In order to have a percolating random graph we should have $k>1$. Here we
have used $k=3$, but some controlling calculations had also been done with $k=4$. In the
actual calculation we have put the RTIM on ER random graphs and study their critical behaviour
by our improved algorithm of the SDRG method. Due to infinite dimensionality
of ER clusters we had to modify some parts of the analysis used in Sec.\ref{sec:234d} for finite D.

\begin{figure}[h!]
\begin{center}
\includegraphics[width=3.4in,angle=0]{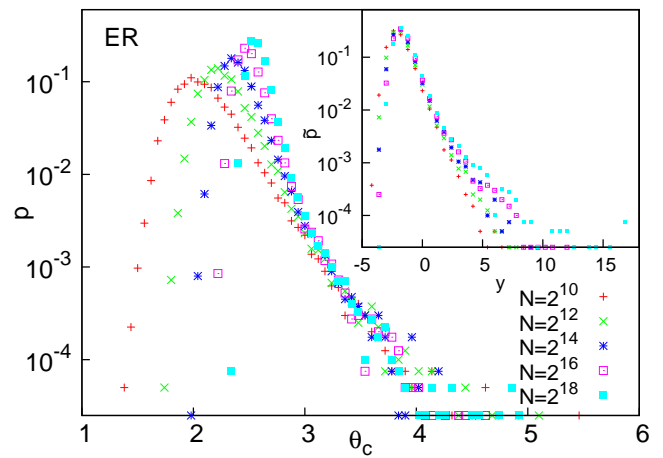}
\end{center}
\vskip -.5cm
\caption{
\label{fig_13} (Color online)
Distribution of the pseudo-critical points, $\theta_c(N)$, for the ER random graphs with box-$h$ randomness.
In the inset the scaled distributions are shown as a function of $y=(\theta_c(N)-\theta_c)N^{1/\omega}$
with $\omega=6$.}
\end{figure}

\begin{figure}[h!]
\begin{center}
\includegraphics[width=3.4in,angle=0]{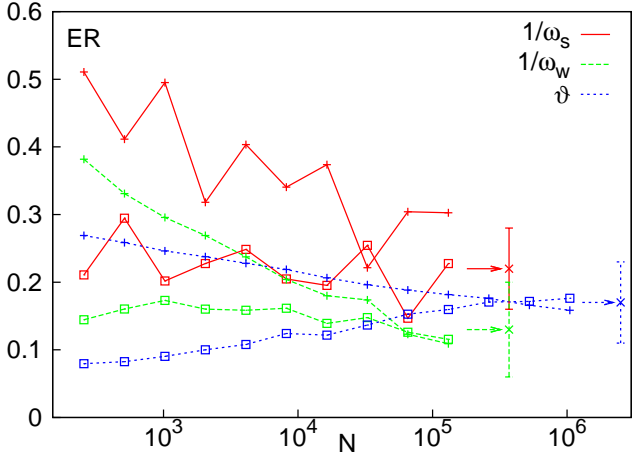}
\end{center}
\vskip -.5cm
\caption{
\label{fig_14} (Color online)
Finite-size estimates for the critical exponents in ER random graphs for the two different disorders (fixed-$h$ +, box-$h$ $\boxdot$). The estimated values for large-$N$ are indicated at the right part of the figure.}
\end{figure}

As for $D \le 4$ we have calculated sample dependent pseudo-critical points, but now
in the doubling method the two identical copies of the sample have been connected by $N/2$ random links.
The distribution of the calculated pseudo-critical points for different values of $N$ are shown in Fig.\ref{fig_13}.
The general behaviour of the distributions is similar to that for finite-D, see Fig.\ref{fig_3} for 4D, but in the present
case for large $\theta_c$ values there is an $N$-independent background of the distributions. This background
is probably due to the large number of connecting random links between the replicas.
This background, however, has a very small weight to the distributions and does not influence
the analysis of the properties of the pseudo-critical points.
Concerning Fig.\ref{fig_13} we have measured the shift,
$\left|\theta_c-\overline{\theta_c}(N)\right| \sim N^{-1/\omega_s}$ and the width $\Delta \theta_c(N) \sim N^{-1/\omega_w}$
of the distributions, in analogy with the finite dimensional problem.
(Compare with Eqs.(\ref{nu_s}) and (\ref{nu_w}), as well as with $\nu_s \to \omega_s/d$ and $\nu_w \to \omega_w/d$, respectively.)
From two-point fits we have calculated effective exponents (see Fig.\ref{fig_14}) from which we have obtained the estimates, $\omega_s=4.5(1.5)$ and $\omega_w=7.8(2.0)$, which are valid for both type of randomnesses. We note that the relative error of the estimates
is somewhat larger than for finite D, but still the two exponents of the distribution agree with each other giving $\omega=6.(2)$.
Using the scaled variable, $y=(\theta_c(N)-\theta_c)N^{1/\omega}$, the distributions show a scaling collapse
as illustrated in the inset of Fig.\ref{fig_13}.

We have studied the fractal properties of the correlation cluster, the average mass of which is found to scale at
the critical point as: $\overline{\mu}(N) \sim N^{\vartheta}$. From two-point fit we have obtained
effective values for $\vartheta$, which are shown in Fig.\ref{fig_14} and which are extrapolated to $\vartheta=0.17(5)$.
We note that according to scaling theory the magnetization exponent of the RTIM on the ER random graph
is given by: $\beta=\omega (1-\vartheta)=5.(2)$.

\begin{figure}[h!]
\begin{center}
\includegraphics[width=3.4in,angle=0]{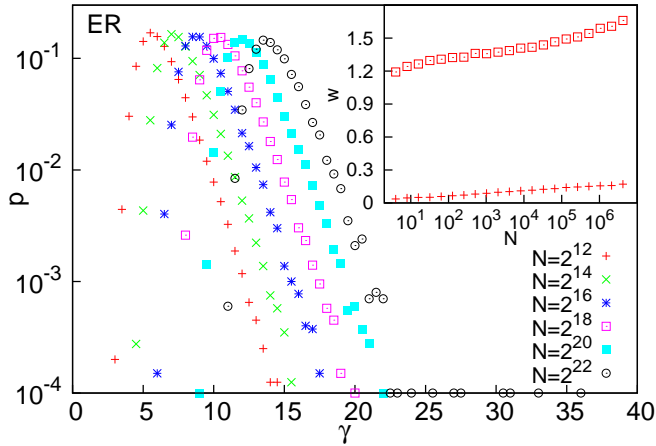}
\end{center}
\vskip -.5cm
\caption{
\label{fig_15} (Color online)
Distribution of the log-energy parameters as a function of the size of the ER clusters for box-$h$ randomness
at the critical point. In the inset the width of the distribution is shown as a function of $\log N$ for
the two different disorders (fixed-$h$ +, box-$h$ $\boxdot$).}
\end{figure} 

We have also investigated the distribution of the log-energy parameters at the critical point, which is shown
in Fig.\ref{fig_15} for different sizes of the ER clusters. In order to see the possible existence
of infinite disorder scaling we have measured the width of the distributions, which are shown in the inset
of Fig.\ref{fig_15} as a function of $\log N$. As seen in the inset the width of the distribution can
be parametrized as $W_0+W_1\log^{\varepsilon} N$, where the constant is $W_0\approx 0$ for fix-$h$ randomness
and it is $W_0 \approx 1.2$ for box-$h$ randomness. In both cases the exponent in the logarithm can be estimated as:
$\varepsilon=1.3(2)$, thus the increase of the width of the distribution is somewhat larger than linear
in $\log N$. This fact justifies that even for ER random graphs the critical behaviour of the RTIM is
controlled by a (logarithmically) infinite disorder fixed point\cite{CP}. Thus we conclude, that the upper critical dimension
of infinite disorder scaling of the RTIM is $D_u=\infty$.
For ER random graphs the singularities of the susceptibility and the specific heat at small temperature
are given by: $\log[T\chi(T)] \sim (2\vartheta-1)|\log T|^{1/\varepsilon}$ and $log[C_V(T)] \sim |\log T|^{1/\varepsilon}$.
\section{Discussion}
\label{sec:discussion}
In this paper we have considered the random transverse-field Ising model, which is a basic model of random
quantum magnets and studied its critical behavior in different dimensions by a variant of the SDRG method.
These investigations are made possible that we have developed an improved numerical algorithm for the SDRG method
so that we could renormalize clusters with up to $N \sim 4 \times 10^6$ sites irrespective of their dimensionality and
topology. We have found strong numerical evidence that the critical behaviour of the RTIM for all dimensions
up to $D=4$ is governed by infinite disorder fixed points. This fact justifies the validity of the use of the SDRG
method as well as indicates that the obtained critical properties of the model, which are summarized in Table.\ref{table:1}
are asymptotically exact. This means that with increasing sizes in the calculation the critical exponents approach
their exact value. We have demonstrated by using different disorder distributions in the initial models that
the strong disorder fixed points are universal, the critical parameters do not depend on the actual form of the disorder.
We have also studied the behaviour of the systems in the vicinity of the critical points and
good agreement with scaling considerations are obtained.

We have considered the upper critical dimension of infinite disorder scaling of the RTIM and studied the critical
behaviour of the model on Erd\H os-R\'enyi random graphs by the improved SDRG algorithm. Our results indicate
that even in this, formally infinite dimensional lattice the critical behaviour of the RTIM is governed by
a (logarithmically) infinite disorder fixed point, thus the upper critical dimension is $D_u=\infty$.

Our results presented in this paper are relevant for several other problems, too, since
the IDFP of the RTIM is expected to govern the critical properties of a large class of random systems,
at least for strong enough disorder. These are, among others,
random quantum ferromagnetic systems having a continuous phase transition
at which a discrete symmetry of a non-conserved order parameter is broken. Examples are the quantum Potts
and clock models\cite{senthil} as well as the Ashkin-Teller model\cite{carlon-clock-at}. Also the quantum spin glass
(QSG) problem could be related to the IDFP of the RTIM. For
the QSG the distribution of couplings in Eq.(\ref{uniform}) contains antiferromagnetic terms, too, however, at an IDFP
frustration is expected to be irrelevant. Thus the critical exponents in Table.\ref{table:1}, with some
appropriate modifications of the scaling relations in the ordered phase\cite{motrunich00} should hold for the QSG,
at least for strong enough disorder.
Also nonequilibrium
phase transitions in the presence of quenched disorder are expected to belong to the universality class of
the RTIM\cite{hiv} and the random walk in a self-affine random potential\cite{selfaff}
could be related to the RTIM.

Finally we mention that the ideas about
the numerical implementation of the SDRG method in Sec.\ref{sec:improved} can be generalized for another models.
In the Appendix we outline the elements of the improved
SDRG algorithm for the random quantum Potts model\cite{senthil}, as well as for the model of disordered Josephson junctions\cite{josephson}. These
results can be used to investigate the critical behaviour of these systems in higher dimensions.

\appendix
\section{Improved SDRG algorithm for other models}
The SDRG approach has been applied for a series of random quantum and classical problems mainly in one dimension.
In higher dimensions the numerical implementation of the SDRG method for these models has basically the same problems
as the na\"{\i}ve algorithm for the RTIM. In these cases one can try to apply and generalize the concept of our improved
algorithm. Here we present the appropriate RG rules for two random quantum models: for the disordered $q$-state quantum
Potts model and for the disordered quantum rotor model, which is a standard model of granular superconductors
and Josephson arrays.

\subsection{Disordered $q$-state quantum Potts model}
In this model at each lattice site, $i$ (or $j$) there is a $q$-state spin variable: $s_i=1,2,\dots,q$
and the Hamiltonian is given by\cite{senthil}:
\be
{\cal H} =
-\sum_{\langle ij \rangle} J_{ij}\delta{\left(s_i,s_j\right)}-\sum_{i} \frac{h_i}{q}\sum_{k=1}^{q-1} M_l^k\;.
\label{eq:Potts}
\ee
Here the first term represents the interaction between the spins and the second term is a generalized transverse field
where $M_i$ is a spin-flip operator at site $i$: $M_i|s_i\rangle=|s_i+1, \mod q\rangle$. As for the random transverse-field
Ising model, what we recover for $q=2$, the $J_{ij}$ couplings and the $h_i$ transverse fields are random variables.
The SDRG decimation rules are very similar to that of the RTIM as described in Sec.\ref{sec:RG}, which differ only in
an extra factor: $\kappa=2/q$.

\underline{$J$-decimation}: the effective transverse fields are given by: $\tilde{h}=\kappa h_i h_j/J_{ij}$.

\underline{$h$-decimation}: the effective couplings are given by: $\tilde{J}_{jk}=\kappa J_{ji}J_{ik}/h_i$, which should
be supplemented by the maximum rule.

In the improved algorithm in Sec.\ref{sec:improved} the distances and ranges in the log-energy space, see
Eqs.(\ref{eq:d}),(\ref{eq:r}) and (\ref{eq:rupdate}), are extended by a constant: $d_0=\ln{(q/2)}=-\ln \kappa$, which now
read as:
\be
d_{ij} = -\ln{J_{ij}}+\frac{l_i}{2}\ln{h_{i}}+\frac{l_j}{2}\ln{h_{j}}+d_0,
\ee
\be
r_i=-\ln{h_i}+d_0,
\ee
\be
\tilde{r}=r_i+r_j-\delta_{ij}+d_0.
\ee
This generalization works for $2\leq q<\infty$.

\subsection{Disordered Josephson junctions}
Here we consider disordered bosons with an occupation operator, ${\hat{n}_i}$, and a phase-variable, $\varphi_i$,
at site $i$. The system is described by the following Hamiltonian\cite{josephson}:
\be
{\cal H} =
-\sum_{\langle ij \rangle} J_{ij}\cos{\left(\varphi_i-\varphi_j\right)}+\sum_{i}U_i{\hat{n}_i}^2\;,
\label{eq:Josephson}
\ee
with random $J_{ij}$ Josephson couplings and $U_i$ charging energies. The SDRG approach has also been
applied to this model resulting in the following RG rules.

\underline{$U$-decimation:}
If the strongest parameter in the Hamiltonian is a grain charging energy $U_i$, then site $i$ is eliminated
and effective couplings are generated between the nearest neighbours of $i$, say $j$ and $k$. In second-order
perturbation calculation this is given by: $\tilde{J}_{jk}=J_{ji}J_{ik}/U_i$, which have to be supplemented
with the maximum rule.

\underline{$J$-decimation:}
If the strongest coupling in the system is a Josephson coupling, $J_{ij}$, then the two sites
form a composite site having an effective charging energy, ${\tilde{U}_i}$, which does not depend on the value of $J_{ij}$
but given by: $\dfrac{1}{\tilde{U}_i}=\dfrac{1}{U_i}+\dfrac{1}{U_j}$.

In the improved SDRG algorithm the distance and the range in Eqs.(\ref{eq:d}) and (\ref{eq:r})
are modified with the substitution $h_i \to U_i$ as:
\be
d_{ij} = -\ln{J_{ij}}+\frac{l_i}{2}\ln{U_{i}}+\frac{l_j}{2}\ln{U_{j}},
\ee
\be
r_i=-\ln{U_i}.
\ee
On the contrary the updated range in Eq.(\ref{eq:rupdate}) has a different form and given by:
\be
\tilde{r}=\ln{\left[\exp{(r_i)}+\exp{(r_j)}\right]}.
\ee

\begin{acknowledgments}
This work has been dedicated to Prof. J\"urgen Hafner on the occasion of his 65th anniversary.
F.I. would like to thank him for the warm hospitality he extended to him during a postdoc
stay in his group in Vienna at 1985-86.
This work has been supported by the Hungarian National Research Fund under grant No OTKA
K62588, K75324 and K77629 and by a German-Hungarian exchange program (DFG-MTA).
We are grateful to D. Huse for helpful correspondence and suggestions and to P. Sz\'epfalusy and H. Rieger for useful discussions.
\end{acknowledgments}


\begin{thebibliography}{99}
\bibitem{spanning_tree} The opposite case, in which only coupling decimations occur, is equivalent to an other well known graph theoretic problem, the minimum spanning tree problem with link weights $w_{ij}=-\ln{J_{ij}}$. In this case the sites are fused together, and the value of the effective transverse field is given by $\ln{h}=W+\sum{\ln{h_i}}$, where $W$ is the total weight of the minimum spanning tree.
\bibitem{subgraph} For correlation clusters the connected subgraphs have spanning property, whereas for any other
clusters the connected subgraphs are non-spanning.
\end{thebibliography}
\end{document}